\documentclass[useAMS,usenatbib]{mn2e}

\usepackage{amsmath}
\usepackage{amssymb}
\usepackage{graphicx}
\usepackage{natbib}
\usepackage{lastpage}

\newcommand{\be}{\begin{equation}}
\newcommand{\ee}{\end{equation}}
\newcommand{\bea}{\begin{eqnarray}}
\newcommand{\eea}{\end{eqnarray}}
\newcommand{\appgeq}{\stackrel{>}{\sim}}
\newcommand{\appleq}{\stackrel{<}{\sim}}

\newcommand{\vbd}{\Delta v_{z,{\rm bend}}}
\newcommand{\vbr}{\Delta v_{z,{\rm breathe}}}
\newcommand{\mnras}{MNRAS}
\newcommand{\apj}{ApJ}
\newcommand{\nat}{Nature}
\newcommand{\aj}{AJ}
\newcommand{\apjl}{ApJL}
\newcommand{\aap}{A\&A}
\newcommand{\aaps}{A\&AS}
\newcommand{\apss}{Ap\&SS}
\newcommand{\pasj}{PASJ}

\title{Bending and Breathing Modes of the Galactic Disk}
\author[L. M. Widrow, J. Barder, M. H. Chequers and E. Cheng]{
Lawrence M. Widrow\thanks{E-mail:
widrow@astro.queensu.ca},
Jarrett Barber,
Matthew H. Chequers and
Edward Cheng\\
Department of Physics, Engineering Physics, and Astronomy, Queen's University, Kingston, ON, K7L 3N6, Canada}

\begin{document}

\date{in original form 2014 January 11}


\maketitle

\label{firstpage}

\begin{abstract}
  We explore the hypothesis that a passing satellite or dark matter
  subhalo has excited coherent oscillations of the Milky Way's stellar
  disk in the direction perpendicular to the Galactic midplane.  This
  work is motivated by recent observations of spatially dependent
  bulk vertical motions within $\sim 2$ kpc of the Sun.  A satellite
  can transfer a fraction of its orbital energy to the disk stars as
  it plunges through the Galactic midplane thereby heating and
  thickening the disk.  Bulk motions arise during the early stages of
  such an event when the disk is still in an unrelaxed state.  We
  present simple toy-model calculations and simulations of
  disk-satellite interactions, which show that the response of the
  disk depends on the relative velocity of the satellite.  When the
  component of the satellite's velocity perpendicular to the disk is
  small compared with that of the stars, the perturbation is
  predominantly a bending mode.  Conversely, breathing and higher
  order modes are excited when the vertical velocity of the satellite
  is larger than that of the stars.  We argue that the compression and
  rarefaction motions seen in three different surveys are in fact
  breathing mode perturbations of the Galactic disk.

\end{abstract}



\section{INTRODUCTION}

Recently, three independent surveys of stellar kinematics within $\sim
2\,{\rm kpc}$ of the Sun detected spatially dependent bulk motions in
the direction perpendicular to the Galactic plane \citep{widrow2012,
  williams2013, carlin2013}.  \citet{widrow2012} found that the bulk
motions of stars, when plotted as a function of position $z$ relative
to the Galactic midplane, have characteristics of a breathing mode
perturbation with a velocity gradient of $\sim 3-5\,{\rm
  km\,s}^{-1}\,{\rm kpc}^{-1}$.  This result was based on a sample of
11K main sequence stars from the Sloan Extension for Galactic
Understanding and Exploration (SEGUE) survey, which focused on
intermediate latitudes above and below the Galactic midplane and
Galactic longitudes in the range $100^\circ < l < 180^\circ$
\citep{yanny2009}.  \citet{williams2013} used a sample of 72K
red-clump stars from the Radial Velocity Experiment (RAVE) survey
\citep{steinmetz2006} to map out bulk motions as a function of
Galactocentric radius and $z$ and found evidence for compressional
motion outside the solar circle and rarefaction inside with peak
vertical bulk velocities of $\pm 15\,{\rm km\,s}^{-1}$.  
\citet{carlin2013} found similar features in their analysis
of 400K F-type stars with proper motions from the PPMXL catalog
\citep{roeser2010} and spectroscopic radial velocities from the
LAMOST/LEGUE survey \citep{cui2012, zhao2012}.  As stressed by
\citet{carlin2013}, the three surveys look in different parts of the
extended solar neighborhood and consider different spatial projections
of the data.

\citet{widrow2012} also found a North-South asymmetry in the number
counts when plotted against $z$.  The number count asymmetry was
confirmed by \citet{yanny2013} who carried out a careful analysis of
the uncertainties and potential systematic effects.  The number counts
show a 10\% (North - South)/(North + South) deficit at $|z|\simeq 400
{\rm pc}$ and an excess of about the same magnitude at $|z|\simeq 800
\,{\rm pc}$.

It is possible that the North-South asymmetries in number counts and
bulk vertical motions are the result of stellar debris from a tidally
disrupted satellite galaxy that is mixing in with the disk stars.
\citet{widrow2012} considered an alternative hypothesis in which the
North-South asymmetries arise from coherent oscillations of the disk
itself, which were excited by a passing satellite or dark matter
subhalo.  Eventually, the oscillations die away due to phase mixing
and Landau damping.  The disk settles into a new equilibrium state,
albeit one with a higher velocity dispersion.  Thus, the bulk motions
seen in the data may indicate an early phase of a disk heating event.
The aim of this paper is to explore this hypothesis in more detail
through toy model calculations and numerical simulations.

A similar question was posed by \citet{minchev2009} in the context of
the stellar velocity distribution in the solar neighborhood.  Detailed
analyses of {\it Hipparcos} data \citep{chereul1998, dehnen1998,
  chereul1999,nordstrom2004} have revealed rich substructure in the
local velocity distribution.  There are classical moving groups, which
are thought to be the stellar streams from dissolved star clusters
(see \citet{eggen1996} and references therein).  Velocity-space
features may arise from dynamical effects of the bar
\citep{dehnen2000} or spiral structure
\citep{desimone2004,quillen2005,chakrabarty2007} or they may be due to
streams of stars that were tidally stripped from accreted satellite
galaxies \citep{navarro2004, helmi2006}.  \citet{minchev2009} consider
a different scenario in which the ``energy kick'' from a passing
satellite leaves ripples in the (disk-plane) velocity distribution of
stars.  Their conjecture is that velocity-space substructure in the
solar neighborhood is a manifestation of these ripples.

Satellites, dark matter subhalos, and globular clusters for that
matter, have long been recognized as possible culprits of disk heating
and thickening.  In general, a massive object that passes through the
disk will transfer a fraction of its orbital energy to the disk stars
\citep{lacey1985,toth1992, sellwood1998}.  Satellite interactions can
also cause the disk to spread out radially and develop warps and
flares \citep{quinn1986,quinn1993,walker1996,velazquez1999}

Satellite encounters can excite various modes in the disk such as
bending modes and breathing modes
\citep{toomre1966,araki1985,mathur1990,weinberg1991}.  It is the
latter that corresponds most closely to the velocity perturbations
seen in the SEGUE, RAVE, and LAMOST surveys.  We will show that a
bending mode perturbation arises when the satellite's vertical
velocity is less than that of the disk stars while breathing and
higher-order modes are excited when the vertical velocity of the
satellite exceeds that of the stars.

Satellites and subhalos can also excite spiral structure and bars in
stellar disks.  (See \citet{sellwood2013} for a recent review of the
stellar dynamics of disk galaxies.)  The seminal work of
\citet{toomre1972} showed that the tidal interaction between a stellar
disk and a companion galaxy of comparable mass can generate grand
design spiral structure similar to what is seen in M51.
Alternatively, multi-armed and flocculent spiral structure can arise
from the continual interactions between the disk and a system of
satellite galaxies and dark matter subhalos.  Cosmological simulations
of structure formation in a $\Lambda$CDM universe suggest that the
halos of Milky Way size galaxies harbour a wealth of substructure in
the form of subhalos \citep{klypin1999,moore1999,gao2004}.  These
results motivated \citet{gauthier2006} and \citet{dubinski2008} to
explore satellite-disk interactions for an M31-like galaxy.  For their
particular M31 model, when the halo is smooth the disk remains stable
against bar formation for 10 Gyr and relatively weak spiral structure
develops, presumably seeded by the shot noise of the N-body
realization.  Conversely, when $\sim 10\%$ of the halo mass initially
resides in compact subhalos, the disk develops prominent spiral
features and forms a strong bar.  Similar results were found in a
series of simulations by \citet{kazantzidis2008}.

More recently, \citet{purcell2011} considered a model Milky Way disk
that was perturbed by a single satellite galaxy.  The prototype for
their perturber was the Sagittarius dwarf spheroidal galaxy
\citep{ibata1994, ibata1997}, which is believed to have survived
several orbits about the Galaxy.  In the \citet{purcell2011}
simulations spiral structure emerges that is similar to the spiral
structure observed in the Milky Way.  \citet{gomez2013} reanalysed
these simulations and found that vertical perturbations in the number
density of disk particles at roughly the Sun's position from the
Galaxy's center were also generated and qualitatively similar to those
seen in \citet{widrow2012} and \citet{yanny2013}.

It is not at all surprising that a Sagittarius-like dwarf produces
vertical oscillations similar to what is found in the data.  The solar
neighborhood is characterized by a circular speed about the Galactic
center of $\simeq 220-230\,{\rm km\,s}^{-1}$ (see, for example,
\citet{bovy2012c} and references therein) and a stellar surface
density of $\simeq 50\,M_\odot\,{\rm pc}^{-2}$ (see, for example,
\citet{holmberg2004}).  Stars in the disk have vertical velocities in
the range of $10-40\,{\rm km\,s}^{-1}$ \citep{robin2003,bovy2012a}.  A
satellite with a comparable surface density to that of the solar
neighborhood, with an orbit that is matched to the local standard of
rest (LSR), and with a vertical velocity through the midplane in
resonance with the vertical motions of disk stars will produce the
strongest perturbations.

In Section 2, we review earlier discussions of disk heating and
thickening and then present a simple toy-model calculation for the
excitation of bending and breathing modes.  In Section 3, we present
results from one-dimensional N-body simulations that support the
toy-model calculation and also illustrate the potentially long-lived
nature of the oscillations.  We provide preliminary results from fully
self-consistent 3D N-body simulations of satellite-disk encounters in
Section 4.  Finally, we summarize our results and give concluding remarks in
Section 5.

\section{BENDING AND BREATHING MODES}

This section focuses on the physics of bending and breathing mode
perturbations of a stellar disk.  We begin with a heuristic discussion
that makes contact with the previous work of disk heating and
thickening.

\subsection{Free-particle approximation}

In an early paper on disk heating, \citet{toth1992} calculated the
transfer of energy to a stellar disk from a passing satellite by
treating the disk stars as free particles that are scattered in the
Keplerian potential of the satellite.  The satellite loses energy via
dynamical friction \citep{chandra1943} and it is this energy that
heats the disk.

Consider a satellite of mass $M_s$ that passes through the disk with
speed $v_s$ at an angle $\theta_s$ relative to the disk normal as
measured in the LSR.  For illustrative purposes, we focus on stars
initially in the plane that contains the satellite's orbit and the
disk normal.  Following \citet{toth1992}, we ignore the random motion of
the stars with respect to the LSR.  The impact parameter for a
star-satellite scattering event is $b = x\cos{\theta_s} +
z\sin{\theta_s}$ where the $\left (x,\,z\right )$ is the position of
the star relative to the point at which the satellite passes through
the Galactic midplane.  The change in the vertical component of the
star's velocity is then
\begin{equation}\label{eq:tothostriker1}
\Delta v_z(x,\,z) =
2v_s \left (1 + \frac{b^2}{b_{90}^2}\right )^{-1}
\left (\cos{\theta_s} + \frac{b}{b_{90}}\sin{\theta_s}\right )
\end{equation}
\noindent where $b_{90}\equiv GM_s/v_s^2$ is the impact parameter
that leads to scattering by $90^\circ$.  Note that
\begin{equation}
b_{90} = 4.3\,{\rm kpc} \left (\frac{M_s}{10^{10}\,M_\odot}\right )
\left (\frac{v_s}{100\,{\rm km\,s^{-1}}}\right )^{-2}~.
\end{equation} 
As a measure of bending and breathing mode perturbations we define
\begin{equation}
\vbd \equiv \frac{1}{2}\left (\Delta v_z\left (x,\,h\right ) + \Delta
  v_z\left (x,\,-h\right )\right )
\end{equation}
and
\begin{equation}
\vbr \equiv \Delta v_z\left (x,\,h\right ) - \Delta
  v_z\left (x,\,-h\right )~.
\end{equation}
In the small and large impact parameter limits we have
\begin{equation}\label{eq:bendlimit}
\vbd = \begin{cases} 2v_s\cos{\theta_s} & b\ll b_{90}\\
2v_s\left (b_{90}/x\right )\tan{\theta_s} & b\gg b_{90}\,;\,b\gg h
\end{cases}
\end{equation}
and
\begin{equation}\label{eq:breathelimit}
\vbr = \begin{cases} 4v_s\left (h/b_{90}\right )\sin^2{\theta_s} & b\ll b_{90}\\
-4v_s\left (hb_{90}/x^2\right )\tan^2{\theta_s} & b\gg b_{90}\,;\,b\gg h~.
\end{cases}
\end{equation}
Thus, both breathing and bending modes are excited by a passing satellite.

\subsection{Resonant interaction}

The previous calculations ignore the epicyclic motions of stars in the
vertical direction.  Stars near the Galactic midplane, where the
gravitational potential $\psi$ is approximately quadratic in the
vertical direction, oscillate about the midplane with frequency
\begin{equation}\label{eq:kappaz0}
\kappa_{z}~\equiv~
\left.\left (\frac{\partial^2\psi}{\partial z^2}\right )^{1/2}\right |_{z=0}~.
\end{equation}  
Stars whose orbits take them further from the midplane oscillate with
a frequency $\nu < \kappa_z$.  \citet{sellwood1998} calculated the
change in vertical energy of a star due to the {\it tidal field} from
the passing satellite in the {\it impulse} approximation (see also
\citet{spitzer1958}).  They averaged over satellite directions and found
\begin{equation}\label{eq:resonant}
\frac{\Delta E_z}{E_c} ~=~ 
\frac{4}{3} \left (\frac{GM_s}{b^2\kappa_z}\right )^2 
\frac{L\left (\beta\right )}{v_s^2}~.
\end{equation}
where $\Delta E_z = \Delta v_z^2/2$, $E_c\equiv h^2\kappa_z^2$ is the
characteristic vertical energy for stars in the disk, $\beta \equiv
2\kappa_z b/v_s$ and the function $L$ is unity for $\beta\to 0$ and is
exponentially small for $\beta\to \infty$.

By focusing on the tidal field, \citet{sellwood1998} pick out the
breathing and higher order modes.  When the satellite velocity is
high, $\Delta E_z\propto v_s^{-2}$ as in the free-particle
approximation (Eq.\ref{eq:breathelimit} with $b\gg b_{90}$).  On the
other hand, if the time scale for a satellite to pass through the disk
is long as compared to $\nu^{-1}$, then the interaction between the
satellite and star will be adiabatic and the energy transfer will be
exponentially small.  The energy transfer peaks when $v_s\simeq
h\kappa_z$ (see Figure 1 of \citet{sellwood1998}), that is, when the
satellite is in resonance with the stellar orbit.

Note that the above equations ignore the finite size of the satellite
in calculating the stellar orbits.  The perturbing force due to a real
satellite will be {\it softened} inside a characteristic scale radius
$a_s$.  In general $a_s = \eta GM_s/\sigma_s^2$ where $\sigma_s$ is
the central velocity dispersion of the satellite and $\eta$ is a
constant of order unity.  Then $a/b_{\rm 90} = \eta v_s^2/\sigma_s^2$.
With a softened perturber, $\vbd\to 0$ for $b\to 0$ whereas for a
point mass perturber, $\vbd$ is nonzero and constant for $b\to 0$,
as in Eq.\,\ref{eq:bendlimit}.

\subsection{Energy transfer to disk stars}

The change in vertical energy of a star due to a perturbing vertical
force $F_z$ is
\begin{equation}\label{eq:workenergy}
\Delta E_z = \int_{-\infty}^t dt' F_z(z,\,t') v_z(t')~.
\end{equation}
In both \citet{toth1992} and \citet{sellwood1998}, the change in
vertical energy of a star is approximated as $\Delta E = \frac{1}{2}
\Delta v_z^2$.  In other words, they assume that $v_z(t)$ in
Eq.\,\ref{eq:workenergy} is generated entirely by $F_z$.  The function
$L(\beta)$ introduced by \citet{sellwood1998} is a crude way of
accounting for the orbital motion.  If the time scale for the
satellite to pass through the disk is long as compared to the orbital
period, then the integral on the right hand side of
Eq.\,\ref{eq:workenergy} approaches zero.

Our interest here is in the details of the vertical perturbations that
are excited by a passing satellite; what modes are excited by a
passing satellite and what is their subsequent evolution?  In the
sections that follow, we address these questions numerically.  Here,
we discuss the excitation of vertical perturbations under the
assumption that the perturbations are small.  We can then approximate
Eq.\,\ref{eq:workenergy} by substituting the unperturbed orbit for
$v_{z0}$ for $v_z$ inside the integral.  This prescription resembles
the Born approximation that is used to calculate scattering amplitudes
in quantum mechanics.

The isothermal plane \citep{spitzer1942,camm1950} provides a simple
equilibrium model for the vertical phase space structure of a stellar
disk.  In this model, the distribution function (DF) is
\begin{equation}\label{eq:spitzer-1} 
f\left (z,\,v_z\right ) = \frac{\rho_0}{\left (2\pi\sigma^2\right
 )^{1/2}}  
e^{-E_z/\sigma^2}
\end{equation} 
where $E_z = v_z^2/2 + \psi(z)$ is the vertical energy and $\sigma$
is the velocity dispersion in the vertical direction.  The density
and potential are given by
\begin{equation}\label{eq:spitzer-2}
\rho(z) = \rho_0\,{\rm sech}^{2}\left (z/z_0\right )
\end{equation}
and
\begin{equation}\label{eq:spitzer-3}
\psi(z) = 2\sigma^2\ln\cosh\left (z/z_0\right )
\end{equation}
where $\rho_0$ is the density at the midplane and $z_0 = \left
  (\sigma^2/2\pi G\rho_0\right )^{1/2}$.  Stars with $E_z\ll \sigma^2$
execute simple harmonic motion about the midplane with a maximum
excursion of $z_{\rm max} \simeq z_0\left (E_z/\sigma^2\right )^{1/2}$
and a period $T \simeq 2\pi/\kappa_{z} \simeq 2^{1/2}\pi z_0/\sigma$
that is approximately independent of $E_z$.  On the other hand, for
stars with $E_z\gg \sigma^2$, we have $z_{\rm max} \simeq z_0\left
  (E_z/2\sigma^2\right )$ and $T \simeq T(E_z) = \left
  (z_0/\sigma\right )\left (8E_z/\sigma^2\right )^{1/2}$.  In what
follows, we set $\sigma$ and $G$ to unity and $z_0=2^{1/2}$ so that
$\rho_0 = 1/4\pi$ and $\kappa_{z}=1$.

\begin{figure}
\includegraphics[width=0.45\textwidth]{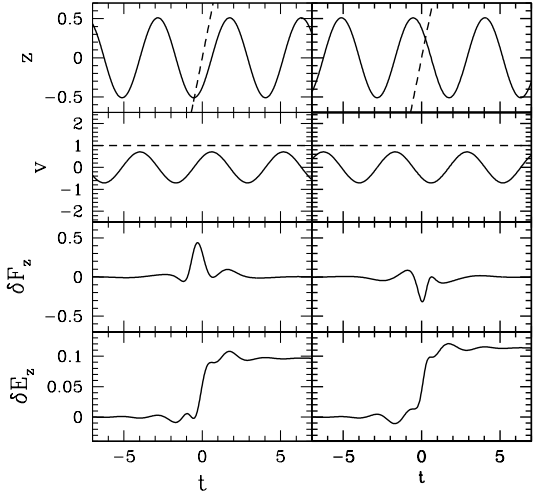}
\caption{Interaction of two stars with a passing satellite.  The left
  and right columns correspond to two stars that differ by $\pi$ in
  their orbital angle $\theta$.  The speed of the satellite is greater
  than the maximum speed of the stars.  The top two rows show the
  phase space coordinates $z$ and $v$ as a function of time for the
  star (solid curves) and satellite (dashed curves).  The third panels
  show the tidal forces acting on the stars while the bottom panels
  show the change in energy $\delta E_z$.}
\label{fig:work1}
\end{figure}

It is useful to introduce the orbital phase angle $\theta$, which is
related to time by the expression $d\theta = 2\pi dt/T(E_z)$.  In Figure
\ref{fig:work1}, we follow the vertical motions of two stars that have
the same $E_z$ but whose phase angles differ by $\pi$.  A satellite
with speed greater than the maximum speed of the stars passes through
the disk at $t=0$.  The upper two panels show the vertical position
and velocity respectively.  The third panel shows the tidal force of
the satellite on the star, that is, the force of the satellite on the
star minus the force of the satellite on a star at $z=0$.  Note that
we've softened the force law with $a_s=1$.  The fourth panel shows the
change in vertical energy of the star,
\begin{equation}\label{eq:deltaenergy}
\delta E_z = \int_{-\infty}^t dt' \left (F_z\left (z,\,t'\right )-F_z\left
(0,\,t'\right )\right ) v_z(t')~.
\end{equation}
Evidently, both stars gain energy.  Likewise, two stars whose phase
angles differ from these two stars by $\pm \pi/2$ will lose energy.
The predominant perturbation in this case is a breathing mode.

\begin{figure}
\includegraphics[width=0.45\textwidth]{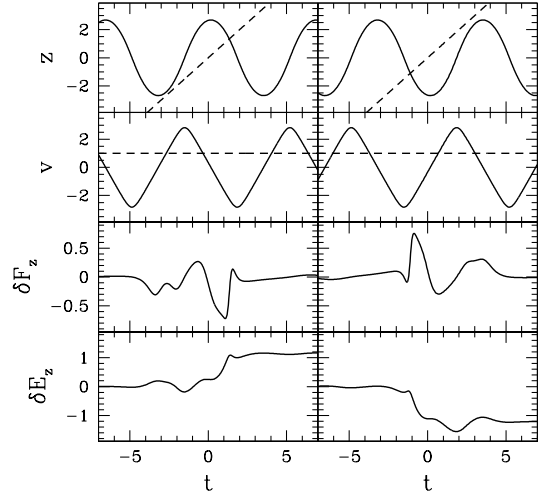}
\caption{Same as Figure \ref{fig:work1} but for the case where the
satellite speed is less than the maximum speed of the stars.}
\label{fig:work2}
\end{figure}

Figure \ref{fig:work2} follows two stars that interact with a
satellite whose speed is less than their maximum speed.  Once again,
the two stars differ in phase angle by $\pi$.  However, in this case,
one star gains energy while the other loses energy.  The perturbation
in this case is predominantly a bending mode.

\begin{figure}
\includegraphics[width=0.45\textwidth]{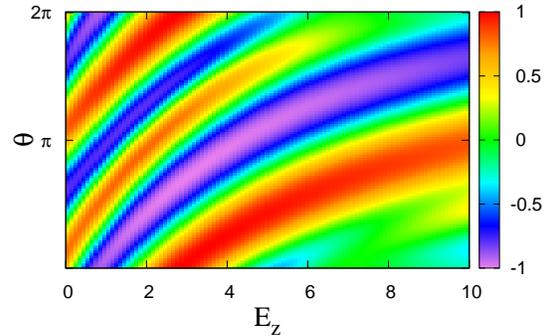}
\caption{Fractional change in vertical energy, $\delta E_z/E_z$, as a
  function of $E_z$ and $\theta$ for $v_s = 1.75$.  The physical value of
  $\delta E_z/E_z$ depends on the relative surface density of the
  perturber and the ``disk''.  Here, we normalize $\delta E_z/E_z$ so
  that its maximum amplitude is unity.}
\label{fig:toydeltaeV1}
\end{figure}

To further study the nature  of the perturbation, we consider the 
Fourier transform of $\delta E_z$:
\begin{equation}
\delta E_z\left (E_z,\,\theta\right ) = \sum_m e^{im\theta} \delta
E_m\left (E_z\right )
\end{equation}
and the corresponding inverse transform
\begin{equation}
\delta E_m\left (E_z\right ) = \frac{1}{2\pi}\int_0^{2\pi} 
d\theta \,e^{-im\theta}\,\delta E\left (E_z,\,\theta\right )~.
\end{equation}
The $m=1$ term corresponds to a bending mode perturbation while the
$m=2$ term corresponds to the breathing mode.  In Figure
\ref{fig:toydeltaeV1} we show the fractional change in energy, $\delta
E_z/E_z$, as a function of $E_z$ and $\theta$ for $v_s=1.75$.  We see
that for $E_z\appleq 5$, where the characteristic orbital speed of the stars
is less than the speed of the satellite, $\delta E_z$ has a pattern
characteristic of a breathing mode (i.e., strong $m=2$ pattern in
$\delta E_z/E_z$  as a function of $\theta$ at fixed energy).  For
$E_z\appgeq 5$, the pattern for $\delta E_z$ is that of a bending mode
($m=1$ pattern).

\begin{figure}
\includegraphics[width=0.45\textwidth]{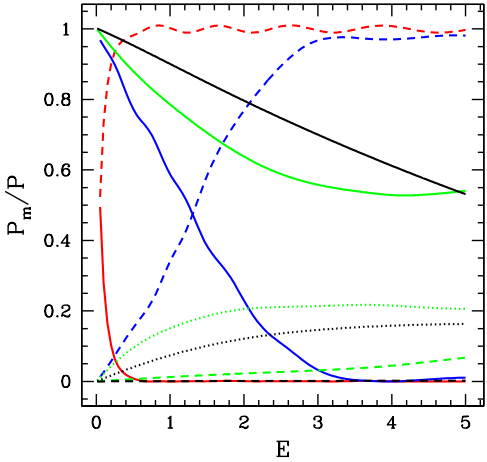}
\caption{Normalized power $P_m/P$ for bending ($m=1$), breathing
  ($m=2$) modes.  Satellite velocities are $v_s = 0.5$ (red), $1$
  (blue), $2$ (green), and $5$ (black).  Bending modes are shown as
  dashed curves while breathing modes are shown as solid curves.  The
  $m=3$ mode is shown for $v_s = 2$ and $5$ as a dotted curve.  (This
  mode is insignificant for smaller values of $v_s$.)  }
\label{fig:power}
\end{figure}

For stars of energy $E_z$ the power in the $m^{th}$ mode of the
perturbation is
\begin{equation}
P_m\left (E_z\right ) = \left |\delta E_m(\left (E_z\right )\right |^2
\end{equation}
where, by Parseval's theorem, the total power is
\begin{equation}
P\left (E_z\right ) = \sum_m P_m\left (E_z\right ) = 
\frac{1}{2\pi}\int_0^{2\pi} d\theta \left (\delta E\left (E_z,\,\theta\right
  )\right )^2~.
\end{equation}
In Figure \ref{fig:power}, we plot the normalized power $P_m/P$ as a
function of $E_z$ for various choices of $m$ and $v_s$.  For our
choice of constants, a star with $E_z=1$ has a maximum orbital speed of
$v_{\rm max} = 1$.  We see that for a slow moving satellite ($v_s=0.5$), the
perturbation is almost entirely a bending mode for $E_z\appgeq 0.4$.  On
the other hand, for a fast moving satellite ($v_s = 2$ or $v_s=5$),
breathing and higher-order modes are excited, especially for
low-energy stars.

\subsection{Toward a realistic model}

The results presented in the previous figures are for a dimensionless
version of the single-component self-gravitating isothermal plane.  By
contrast, the stellar disk of the Milky Way is better described by
multiple components and the Galactic potential has contributions
from the bulge and dark halo as well as the disk.

To make contact with the actual Milky Way, we consider several
observationally motivated models.  \citet{kuijken1989} modeled the DF
of stars and the vertical potential in the extended solar neighborhood
based on a survey of $\sim 500$ K dwarfs toward the south Galactic
pole.  They chose the following parametric form for the potential:
\begin{equation}\label{eq:kgpotential}
\psi(z) = K\left (\left (z^2 + D^2\right )^{1/2}-D\right ) +
  Hz^2~.
\end{equation}
The first term on the right hand side is meant to account for the
disk's contribution to the potential where $D$ is the effective
thickness of the disk and $K$ is proportional to the disk's surface
density.  The second term is meant to account for the halo where $H$
is proportional to the effective halo density in the solar neighborhood.
\citet{kuijken1989} fixed $D$ to be $180\,{\rm pc}$ and imposed the constraint
\begin{equation} 
H = 2\pi G\left (0.015 - 9.4\times 10^{-5}\left (\frac{K}{M_\odot {\rm
       pc}^{-2}}\right )\pm 0.002\right ) M_\odot {\rm pc}^{-3}
\end{equation} 
to be consistent with observations of the Galactic rotation curve.
They found
\begin{equation}
K = 2\pi G\left (46\pm 9\,M_\odot\,{\rm pc}^{-2}\right )
\end{equation}
and therefore
\begin{equation}
H = 2\pi G\left (0.011\pm 0.003 \, M_\odot {\rm pc}^{-3}\right )~.
\end{equation}

\begin{table*}
\begin{tabular}{ccccc}
 \hline
  Model   & D (pc) & $K/2\pi G ~\left (M_\odot\,pc^{-2}\right )$ 
 & $H/2\pi G ~\left (M_\odot\,pc^{-3}\right )$ & $\kappa_{z}~\left
   (Myr\right )$  \\
\hline
KG & 180 & $46\pm 9$ & $0.011\pm 0.003$ & $8.5\pm 1.7$\\
Besan\c{c}on & 320 & 33 & 0.019 & 6.1 \\
WPD & 540 & 58 & 0.006 & 5.6\\
\hline
\end{tabular}
\caption{Parameters for the vertical potential fitting formula
  Eq.\,\ref{eq:kgpotential}.  Models are \citet{kuijken1989} (KG),
  \citet{robin2003} (Besan\c{c}on), and the most stable of the
  \citet{widrow2008} (WPD) Milky Way models.  The vertical epicyclic
  frequency for stars near the midplane, $\kappa_{z}$ is given by
  Eq.\,\ref{eq:kappaz0}.}
\end{table*}

The Besan\c{c}on model \citep{robin2003}, which is one of
the most widely cited models of the Milky Way, provides a self-consistent
description of the Galactic potential and stellar populations for the
Milky Way based on star counts and the rotation curve.  The model is
presented as a set of density-laws for the (multi-component) thin
disk, the thick disk, the bulge, the stellar halo, and the dark halo.
From these results, it is straightforward to extract the vertical
potential in the solar neighborhood.

Apart from the physical interpretation ascribed by
\citet{kuijken1989}, Eq.\ref{eq:kgpotential} provides a convenient
parametric form for the vertical potential in part because the inverse
$z=z\left (\psi\right )$ is analytic.  With the parameters given in
Table 1, Eq.\,\ref{eq:kgpotential} provides an excellent fit to
$\psi(z)$ for the Besan\c{c}on model in the range $0<z<2$.

\citet{widrow2008} presented a set of twenty-five disk-bulge-halo
models for the Milky Way designed to fit observational data for the
rotation curve and local kinematics.  Each model is represented by
self-consistent expressions for the DF and density of the three
components as well as the total gravitational potential.  The models
are characterized by their susceptibility to bar and spiral
instabilities.  Here, we consider the most stable model (Toomre
parameter $Q=2$,
global stability parameter $X=4.5$) and again fit the vertical
potential at $R=R_{\rm sun}$ to Eq.\,\ref{eq:kgpotential}.

\begin{figure}
\includegraphics[width=0.45\textwidth]{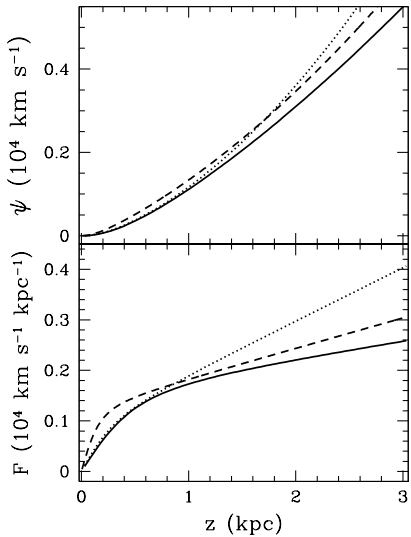}
\caption{Vertical potential (top panel) and force (bottom panel) 
as a function of $z$ for the \citet{kuijken1989} model (dotted
  curve), the Besan\c{c}on model (dashed curve), and the most stable
  \citet{widrow2008} model (solid curve).}
\label{fig:potential}
\end{figure}

\begin{figure}
\includegraphics[width=0.45\textwidth]{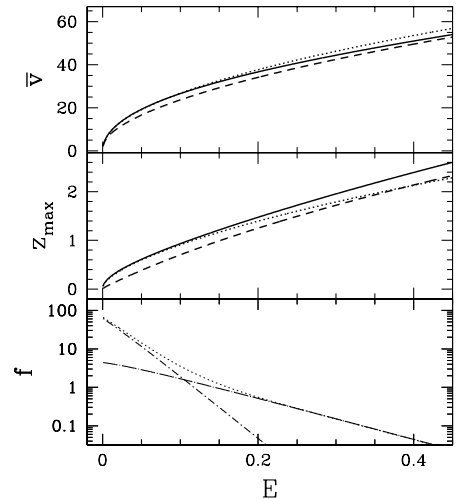}
\caption{Mean speed, $\bar v$, and maximum excursion from the midplane
  $z_{\rm max}$ as a function of $E_z$ for the models considered in
  Section 2.4.  Line types are the same as in Figure
  \ref{fig:potential}.  Units are $100\,{\rm km\,s}^{-1}$, ${\rm
    kpc}$, and $10^4\,\left ({\rm km\,s}^{-1}\right )^2$ for
  $\bar{v}$, $z_{\rm max}$ and $E$ respectively.  Shown in the third
  panel is the DF for the \citet{robin2003} model.  The thin and thick
  disks are shown as dash-dot and long dash-dot lines, respectively.}
\label{fig:tofe}
\end{figure}

The vertical potential and force for the three models are shown in
Figure \ref{fig:potential}.  Figure \ref{fig:tofe} shows the mean
speed and maximum excursion in $z$ for the three models considered
here as well as the DF for the Besan\c{c}on model.  Consider, for
example, stars that have $z_{\rm max}= 1\,{\rm kpc}$.  These stars
have a mean speed of about $25-30\,{\rm km\,s}^{-1}$ with the thick
and thin disks making similar contributions to the total stellar DF.

\section{EVOLUTION OF VERTICAL OSCILLATIONS}

The mode decomposition scheme described above provides a useful
starting point for understanding the evolution of perturbations in a
stellar disk.  Bending, breathing, and higher order modes are
collective excitations \citep{araki1985, sellwood1998, mathur1990,
  weinberg1991}, which can lose energy via Landau damping and
dynamical friction with the halo.  Phase mixing also effects the
evolution of the modes and may lead to an effective damping of
observables, which invariably involve a projection of the
coarse-grained DF.  Finally, the interaction region of the disk for a
satellite encounter will be sheared by differential rotation.  In what
follows, we provide a brief discussion of these effects and then
present one-dimensional numerical simulations of a satellite that
interacts with a plane-symmetric disk.

\subsection{Theoretical considerations}

It is a well known result that perturbations in a homogeneous
self-gravitating fluid with density $\rho_0$ and sound speed $c_s$
grow via the Jeans instability if the wave length of the perturbation
is greater than the Jeans length $\lambda_J \equiv \left (\pi
  c_s^2/G\rho_0\right )$.  Perturbations whose wavelength $\lambda$ is
less than $\lambda_J$ oscillate as sound waves with period $T = 2\pi
\lambda/c_s$ (see for example \citet{binney2008}).  In a homogeneous
collisionless system with Maxwellian DF, the Jeans length is the same
as for a fluid except that the sound speed is replaced by the velocity
dispersion.  However, perturbations with $\lambda < \lambda_J$ are
strongly damped by a process known as Landau damping, which was first
discussed in the context of plasma waves \citep{landau1946}.  
(For an excellent pedagogical discussion, see
\citet{stix1992}).  Particles can draw energy from or feed energy
to the collective mode.  In the case of a spatially homogeneous
Maxwellian DF, the net effect of the stars is to damp the wave on a
time-scale comparable to the period of the oscillations
\citep{lyndenbell1962}.

As stressed in \citet{binney2008} and elsewhere, Landau-damped waves
are not true modes but rather the response of the system to a
perturbation.  True modes of the system would satisfy the
collisionless Boltzmann and Poisson equations with a harmonic
time-dependence for all times.
 
In the case when the unperturbed system is spatially homogeneous, one
must resort to the Jeans swindle, which side-steps the fact that the
unperturbed potential is ill-defined.  For the isothermal plane, the
existence of a self-consistent equilibrium solution
(Eqs.\,\ref{eq:spitzer-1}-\ref{eq:spitzer-3}) allows one to carry out
a perturbation analysis without resorting to the Jeans swindle
\citep{araki1985}.

In principle, wavelike perturbations of the isothermal plane will also
experience Landau damping, which occurs when there are particles in
the distribution in resonance with the wave, i.e., whenever the
frequency of the wave matches $n\nu$ where $\nu$ is the orbital
frequency of a star and $n$ is an integer.  The orbital frequencies
$\nu$ of stars in the isothermal plane range from $0$ to $\nu_{\rm
  max}=\kappa_z$.  Thus, the spectrum of orbits and their harmonics
consist of semi-infinite overlapping segments $\left \{0,n\nu_{\rm
    max}\right \}$ where $n$ is a positive integer.  For a mode of
frequency $\nu$, there will be a set of particles with $n > {\rm
  int}\left (\nu/\nu_{\rm max}\right )$ that are in resonance with and
therefore capable of Landau-damping, the mode.

\citet{mathur1990} and \citet{weinberg1991} showed that true modes,
that is, modes that do not suffer Landau damping, exist for the {\it
  truncated} isothermal plane.  In that model, the distribution
function is given by
\begin{equation}\label{eq:TIP}
f\left (z,\,v_z\right ) = \begin{cases}
f_0\left  (e^{-E_z/\sigma^2}-e^{W/\sigma^2}\right ) & \mbox{for}~ 0 < E_z<W \\
0 & \mbox{otherwise}~.
\end{cases}
\end{equation}
Orbital frequencies range from $\nu_{\rm min}$ to $\nu_{\rm max}$
where $\nu_{\rm min}$ is the orbital frequency of a star with energy
$E_z=W$ and $\nu_{\rm max} = \kappa_z$ is the frequency for a star
with $E_z\to 0$, that is, a star whose orbit stays near the midplane.
The frequency spectrum for orbits has a gap from $0$ to $\nu_{\rm
  min}$, called the principal gap (\citet{mathur1990};
\citet{weinberg1991}).  If $2\nu_{\rm min}>\nu_{\rm max}$, then there
will be a second gap between $\nu_{\rm max}$ and $2\nu_{\rm min}$, and
so on.  \citet{weinberg1991} showed that a mode exists for $\nu = 0$.
This mode corresponds to a displacement of the system as a whole and
corresponds to a bending mode.  Once the system is embedded in an
external potential, the frequency of the mode shifts to a non-zero
value.

\citet{weinberg1991} analyzed the linearized collisionless
Boltzmann and Poisson equations and showed that modes also exist in
the higher frequency gaps and typically lie near the upper freqency
end of the gap.  Thus, for the first gap beyond the principal one, the
frequency of the mode is less than but close to $2\nu_{\rm min}$.
That is, the coherent oscillations are near the $2:1$ resonance with
particles at the upper edge of the energy distribution.
\begin{figure}
\includegraphics[width=0.45\textwidth]{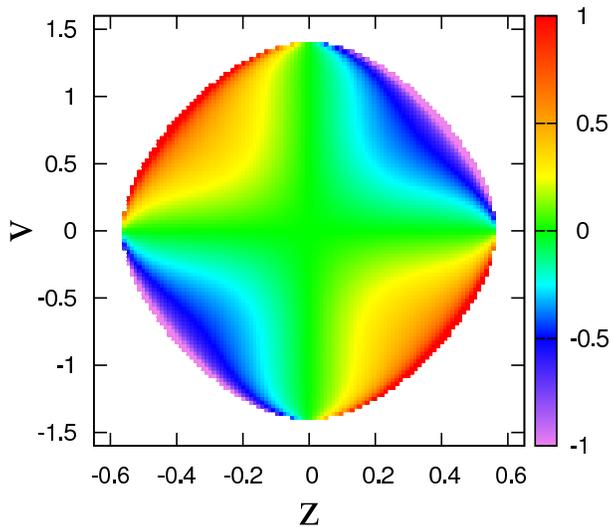}
\caption{Color map of the distribution function perturbation for
the case $W=2\sigma^2$.  The calculation was carried out using the
method described in \citet{weinberg1991}.}
\label{fig:weinberg_DF}
\end{figure}

\begin{figure}
\includegraphics[width=0.45\textwidth]{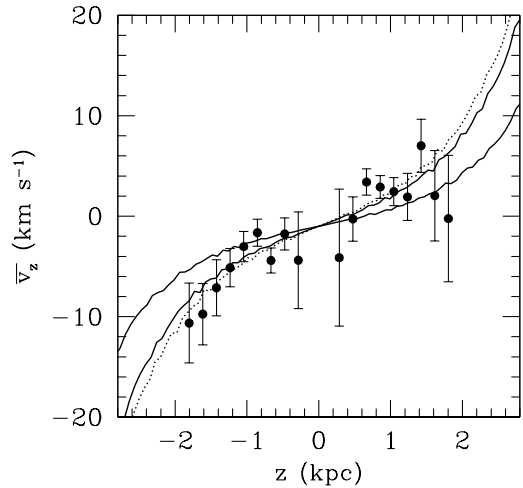}
\caption{Bulk velocity $\bar{v_z}$ as a function of position relative
  to the Galactic midplane for the phase space perturbation shown in
  Figure \ref{fig:weinberg_DF}.  Data points are from
  \citet{widrow2012}.  The dotted curve corresponds to the phase thats
  shown in Figure \ref{fig:weinberg_DF}, where the velocity
  perturbation is at its maximum.  The other two (solid) curves
  correspond to different phases (i.e., rotation of Figure
  \ref{fig:weinberg_DF} by $15^\circ$ or $30^\circ$.}
\label{fig:moments}
\end{figure}

In Weinberg's analysis, the phase space DF is written as the sum of
the zeroth-order (equilibrium) solution and a linear perturbation.
Figure \ref{fig:weinberg_DF} shows the phase space perturbation for
the breathing model of a $W=2\sigma^2$ model.  The perturbation
rotates in the clockwise direction with a period given by the orbital
period of stars near the energy cut-off, $\nu_{\rm min}$ while the
pattern is periodic with a frequency of $2\nu_{\rm min}$.  The phase
shown in Figure \ref{fig:weinberg_DF} is consistent with a velocity
perturbation similar to what is seen in the data.  In the South
($z<0$) there are more stars with negative velocity than positive
velocity.  The situation is the reverse in the North.  Figure
\ref{fig:moments} shows the bulk velocity perturbation (essentially,
the first moment of the DF) for the phase shown in Figure
\ref{fig:weinberg_DF} and for two other phases in the perturbation's
cycle.  The magnitude of the perturbation has been adjusted to fit the
data.  We note that the slope of the bulk velocity profile increases
as one moves away from the midplane.

As noted above, in the absence of an external potential, the bending
mode for this one-dimensional model corresponds to a displacement of
the system in position and velocity.  In an actual stellar disk, there
is a restoring force to a local vertical displacement due to the
gravitational field from the rest of the disk.  One can model this
effect in the one-dimensional model by adding an external potential.
The mode then oscillates but, as discussed in \citet{weinberg1991},
its structure is qualitatively similar to a simple displacement.

Observationally, a bending mode manifests itself as a shift in the
position and velocity of the midplane across the disk.  Indeed,
midplane displacements on the order of $10-100$ pc have been found in
the analysis of CO observations by \citet{nakanishi2006}.  On the
other hand, if we focus on the vertical structure of the disk in the
solar neighborhood, then a bending mode effects our determination of
the Sun's vertical position and velocity but is otherwise
unobservable.

\subsection{Simulations in one dimension}

\citet{widrow2012} presented one-dimensional N-body simulations to illustrate
the evolution of a perturbed isothermal plane.  In these simulations
the DF is sampled by a set of ``particles'' each of which represents
an infinite plane with surface density $\Sigma_p$.  The force acting
on the $i$'th particle is
\begin{equation}\label{eq:1Dsimulation}
F_i = 2\pi G\Sigma_p\left (N^R_i - N^L_i\right )
\end{equation}
where $N^R_i$ ($N^L_i$) is the number of particles to its right
(left).  The equilibrium model was the Spitzer solution while the
perturbation was chosen to correspond to the velocity and number
density perturbations seen in the data.  Waves appear to reflect off
the low density regions above and below the midplane.  The
perturbation also appears to decay from the inside out.  This result
is consistent with the result discussed above (see also
\citet{weinberg1991}, namely that coherent excitations of the system
are mainly a phenomena of stars with higher vertical energies and also
that those high-energy stars are most easily excited by a passing
satellite.

\begin{figure}
\includegraphics[width=0.65\textwidth]{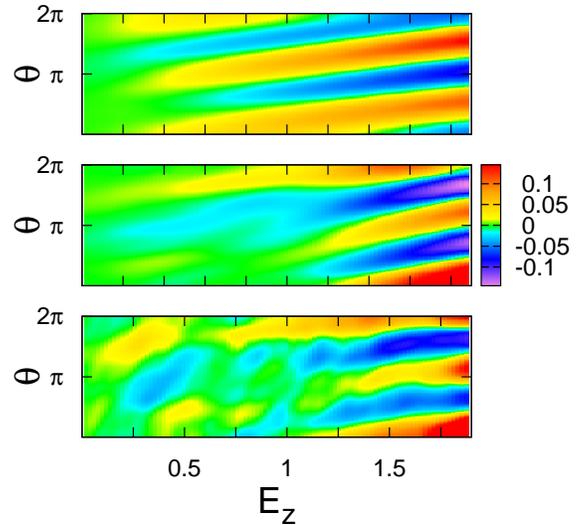}
\caption{Fractional change in vertical energy, $\delta E_z$ as a
  function of $E_z$ and $\theta$ due to a satellite with speed $v_s =
  0.25$.  Top, middle, and bottom panels correspond to $t=0,\,3T$, and
  $30T$ respectively where $T=2\pi$ is the orbital period for a star
  near the midplane.}
\label{fig:1Dsimulation} 
\end{figure}

We have carried out a new simulation where the initial conditions are
that of the truncated isothermal plane (Eq.\,\ref{eq:TIP}) with
$W=2\sigma^2$ (see also \citet{weinberg1991}).  The system is
perturbed by a passing satellite whose velocity is $v_s = 2\sigma$ and
whose surface density is equal to one half the surface density of the
system.  In Figure \ref{fig:1Dsimulation} we show the change in the DF
as a function of $E_z$ and $\theta$ for three epochs. At the first
epoch ($t=0$) the satellite is crossing the midplane.  The second and
third panels show the change in the DF for $t\simeq 3T$ and $t\simeq
30T$ where $T=2\pi$ is the period for stars with $E_z\to 0$.  A
bending-mode perturbation is generated for stars with $E_z\appgeq
0.5$.  As the system evolves, the perturbation is washed out for
$E_z\appleq 1.2$, presumably by a combination of phase mixing and
Landau damping.  However, the perturbation persists mainly for stars with
energies near the energy cut-off.  This result is jibes with the fact
that the true modes of the system are most pronounced near the energy
cut-off.  (See Figure 4 from \citet{weinberg1991} and our Figure
7)

\section{N-BODY SIMULATIONS}

\subsection{Single satellite perturbations}

As discussed above, the coupling between a satellite and the vertical
modes of the Galactic disk in the solar neighborhood is governed by
the relative in-plane motions of the satellite and the LSR, the match
between the satellite's vertical velocity and the vertical epicyclic
motions of solar neighborhood stars, and the mass and concentration of
the satellite.  \citet{gomez2013} used numerical simulations to show
that the Sagittarius dwarf galaxy could have induced wavelike
perturbations as it plunged through the stellar disk.  The initial
Milky Way model is from \citet{widrow2008}.  \citet{gomez2013}
considered two models for the Sagittarius progenitor, a light model,
with virial mass $M_{\rm vir} = 3\times 10^{10}\,M_\odot$, an NFW
\citep{navarro1996} scale length $r_s=4.9\,{\rm kpc}$, and a
concentration parameter $c=16.3$ and a heavy model with $M_{\rm vir} =
10^{11}\,M_\odot$, $r_s = 6.5\,{\rm kpc}$ and $c=18$.  The surface
densities for these models are $220\,{\rm M}_\odot\,{\rm pc}^{-2}$ and
$380\,{\rm M}_\odot\,{\rm pc}^{-2}$.  These values are larger than the
disk surface density in the solar neighborhood.  However the
satellites suffer mass loss due to tidal stripping before the pass
through the disk.  In any case, it is not surprising that the disk is
perturbed by a passing satellite with these parameters and indeed,
\citet{gomez2013} find that there are regions in the disk
characteristic of the solar neighborhood (that is, $\sim 8\,{\rm kpc}$
from the Galaxy's center) where the vertical number density profile of
disk stars has wave-like perturbations qualitatively similar to what
is seen in the data \citep{widrow2012,yanny2013}.
 
\begin{figure}
\includegraphics[width=0.45\textwidth]{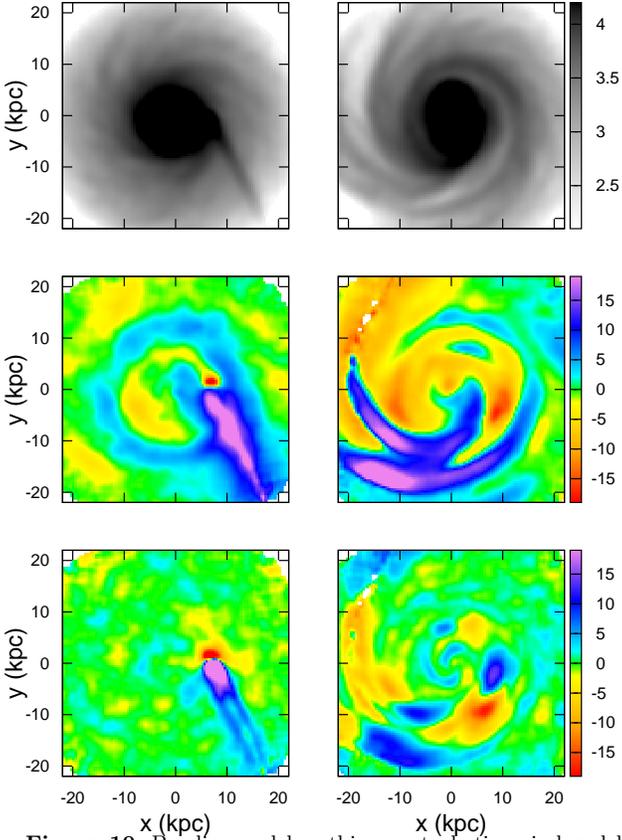}
\caption{Bending and breathing perturbations induced by a $4\times
10^9\,M_\odot$ satellite.  The left column shows the disk just as the
satellite is passing through the midplane while the right column
shows the disk at $t=250\,{\rm Myr}$.  The top panels shows the
density map (logarithmic grey-scale shading).  The middle panels
show the bending mode parameter $B$ as defined in
Eq.\ref{eq:bendbreath}.  The units for $B$ are ${\rm km\,s}^{-1}$.
The bottom set of panels shows the breathing mode parameter $A$ in
units of ${\rm km\,s}^{-1}{\rm kpc}^{-1}$.}
\label{fig:singlesat}
\end{figure}

Here we present preliminary results from simulations of satellite-disk
interactions with a particular focus on vertical velocity
perturbations.  We choose the most stable of the Galactic models
presented in \citet{widrow2008} as our model for the parent galaxy.
In Figure \ref{fig:singlesat} we present results for a satellite that
passes through the midplane of the Milky Way on a prograde orbit at a
Galactocentric radius of $8\,{\rm kpc}$ with a vertical speed of
roughly $60\,{\rm km\,s}^{-1}$.  The satellite has a mass of $M_s =
4\times 10^9\,M_\odot$ and is truncated at a radius of $\sim 0.9\,{\rm
  kpc}$.  The first column of panels shows the disk as the satellite
is passing through the midplane while the second column shows the disk
some $250\,{\rm Myr}$ later.  The top panels show a logarithmic map of
the surface density across the disk.  The strong wake generated by the
satellite is clearly visible in the upper left panel.  Over time, the
disk develops prominent flocculent spiral structure.

To quantify the bending and breathing modes we model the bulk vertical
velocity across the disk plane as
\begin{equation}\label{eq:bendbreath}
  \bar{v}_z\left (x,\,y;\,z\right ) ~=~
  A\left (x,\,y\right )z + B\left (x,\,y\right )~.
\end{equation}
That is, for disk stars within a two-dimensional cylinder centered on
the point $\left (x,\,y\right )$ in the disk plane we fit a
$\bar{v}_z$ to a linear function in $z$.  The coeffient $B$ is a
measure of the bending mode strength and is shown in the two middle
panels of Figure \ref{fig:singlesat} while $A$ is a measure of the
breathing mode strength and is shown in the bottom two panels.  We see
that the satellite excites both modes as it passes through the disk.
These perturbations are sheared by the differential rotation of the
disk and continue to oscillate and reverberate for many 100's of Myr.
Both modes have a pattern across the disk that is qualitatively
similar to the spiral structure seen in the density map.

\subsection{System of satellites}

\citet{gauthier2006} followed the evolution of a disk-bulge-halo
galaxy in which the halo comprises a smooth component with an NFW
profile \citep{navarro1996} and a system of 100 subhalos.  For the
parent galaxy they used the self-consistent equilibrium model of M31
from \citet{widrow2005} labelled M31a.  This model provides a good
match to the observed rotation curve, surface brightness profile, and
velocity dispersion profile and is stable against the formation of a
bar for $10\, {\rm Gyr}$ so long as the halo mass is assumed to be
smoothly distributed.  The disk has a mass of $M_D = 7.8\times
10^{10}\,M_\odot$ and an exponential scale length of $R_D = 5.6\,{\rm
  kpc}$.  The circular speed curve reaches a peak value of $260\,{\rm
  km\,s}^{-1}$ at a radius of $\sim 10\,{\rm kpc}$.  The complete list
of model parameters can be found in Table 2 of \citet{widrow2005} or
Table 1 of \citet{gauthier2006}.  In principle, the model could be
``rescaled'' to make better contact with Milky Way observations.
Nevertheless, the qualitative features of the simulations should be
applicable to the Galaxy.  A proper Milky Way version of this
numerical experiment will be presented in a forthcoming publication.

\begin{figure}
\includegraphics[width=0.45\textwidth]{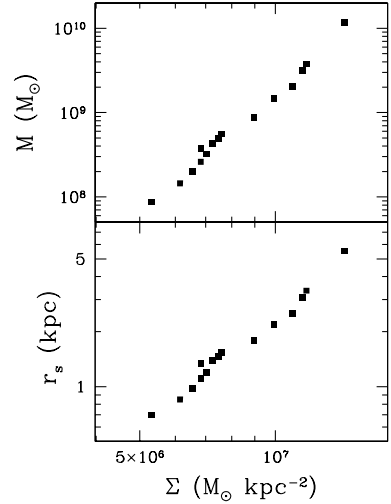}
\caption{Mass and NFW scale radius as a function of average surface
  density within a projected surface density $r_s$ for the subhalos in
  the \citet{gauthier2006} simulation.}
\label{fig:satmass}
\end{figure}

The 100 subhalos in the \citet{gauthier2006} simulation range in mass
from $8.7\times 10^7-1.2\times 10^{10}\,M_\odot$ with a number density
mass function given by $dN/dM_s\propto M_s^{-1.9}$ \citep{gao2004}.
Initially, each subhalo is modelled as a spherically symmetric
truncated NFW system where the truncation radius $r_t$ is given by its
tidal radius, as determined by the Jacobi condition, at $50\,{\rm
  kpc}$:
\begin{equation}
\bar{\rho_s}\left (r_t\right ) = 3\bar{\rho_h}\left (r=50\,{\rm
      kpc}\right )~.
\end{equation}
Here $\bar{\rho}(r)$ is the mean (sub)halo density inside radius $r$.
The radius $50\,{\rm kpc}$ is the mean apocenter for the initial
system of subhalos.  Each subhalo is characterized by its mass $M_s$,
scale radius $r_s$, and concentration $c = r_t/r_s$.  In Figure
\ref{fig:satmass}, we plot $M_s$ and $r_s$ as a function of the
average surface density within a projected radius $R = r_s$,
$\bar{\Sigma}_s(R=r_s)$ for individual subhalos (see Table 2 of
\citet{gauthier2006}).  Note that $\Sigma(R\appleq r_s/10)$ is
approximately constant and a factor of $\simeq 10$ larger than
$\bar{\Sigma}_s(R=r_s)$.
 
\begin{figure}
\includegraphics[width=0.45\textwidth]{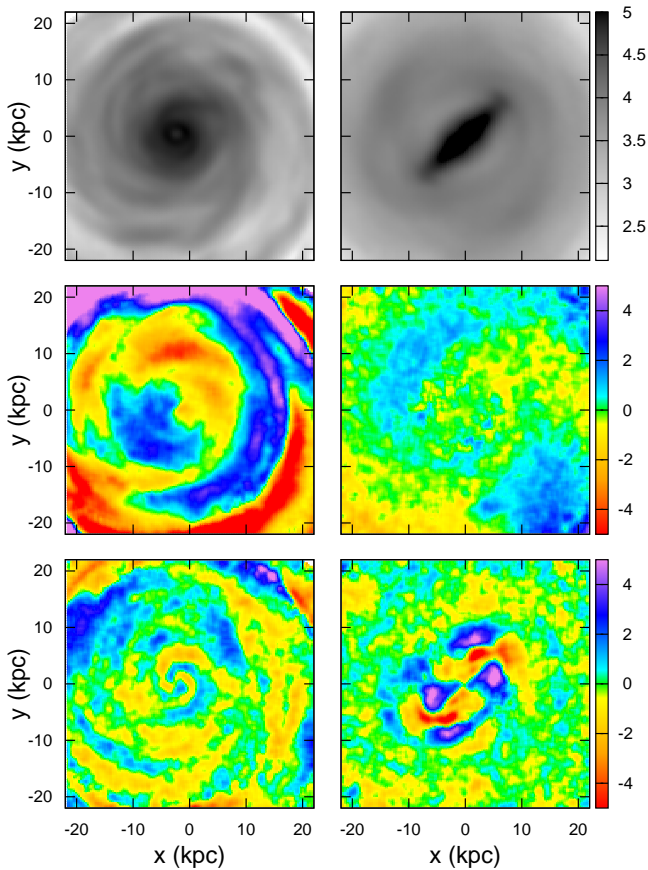}
\caption{Face-on map of the surface density and the breathing and
  breathing mode strengths for the \citet{gauthier2006} simulation.
  Left panels show the galaxy at $2.5\,{\rm Gyr}$, which is prior to
  the formation of the bar while the right panels show the galaxy at
  $10\,{\rm Gyr}$, which is well after the bar has formed.  The top
  panels show the logarithmic grey-scale maps of the density.  The
  middle panels show the bending mode strong in units of ${\rm
    km\,s}^{-1}$ while the bottom panels show the breathing mode
  strength in units of ${\rm km\,s}^{-1}{\rm kpc}^{-1}$.}
\label{fig:gauthier}
\end{figure}

One of the most striking results from the \citet{gauthier2006}
simulation is the formation of a strong bar at about $5\,{\rm Gyr}$
(see, also \citet{kazantzidis2008}).  In Figure \ref{fig:gauthier}, we
show the surface density and breathing mode strength at $2.5\,{\rm
  Gyr}$ and $10\,{\rm Gyr}$.  The length of the bar is $\sim 20\,{\rm
  kpc}$ as can be seen in the upper right panel of Figure
\ref{fig:gauthier} and Figure 5 of \citet{gauthier2006}.  Prominent
spiral structure also develops and indeed, appears to be a precursor
to the formation of the bar.  Though there is some disk heating during
the first $4\,{\rm Gyr}$ the most significant heating and thickening
occurs after the bar forms (see \citet{dubinski2008} for a further
discussion).  By contrast, no bar and only weak spiral structure
develops in the control experiment, which assumes a smooth halo.
Evidently, satellites and subhalos provoke spiral structure and bar
formation (see also \citet{purcell2011}).

It is not at all surprising that system of subhalos also excites
vertical oscillations in the stellar disk. \footnote{See the animation
  found at the URL http://www.cita.utoronto.ca/\~dubinski/Rome2007/.
  Note in particular, the switch half-way through to an edge-on view
  of the disk.}  Bending and breathing mode perturbations are found
across the disk and throughout the simulation, as can be seen in the
middle and lower panels of Figure \ref{fig:gauthier}.  Prior to
  bar formation there are strong large-scale bending modes across the
  disk with amplitudes on the order of $4\,{\rm km\,s}^{-1}$.  The
  breathing modes have a somewhat smaller amplitude ($\sim 2\,{\rm
    km\,s}^{-1}$) and vary on smaller scales.  At later times, after
  the bar has formed, the bending mode perturbations have diminished.
  Moreover, in the inner parts of the Galaxy, the breathing mode
  mirrors the bar.  Thus, while subhalos may have triggered the
  formation of the bar, it is the bar that generates and maintains
  compression and rarefaction motions in the inner Galaxy.  Of course,
  the Sun sits well beyond the region of the bar and it is therefore
  unlikely that the bar could cause the bulk motions seen in the solar
  neighborhood.

\section{DISCUSSION and CONCLUSIONS}

The implications of spatially dependent bulk motions
perpendicular to the Galactic disk were highlighted by
\citet{oort1932} in his seminal work on the structure of the Galactic
disk.  Oort's aim was to determine the potential $\psi(z)$ from the
local stellar density and velocity distribution.  He based his
analysis on the assumption that the local distribution of stars is in
equilibrium.  To test the assumption, he computed the mean vertical
velocity for stars in four separate bins: $100\,{\rm pc} < \pm z <
200\,{\rm pc}$ and $200\,{\rm pc} < \pm z < 500\,{\rm pc}$ but did not
find evidence for systematic motions, a result that he notes ``lends
some support to the assumption \ldots that in the $z$-direction the
stars are thoroughly mixed'' \citep{oort1932}.  Turning Oort's
argument around, the detection of bulk vertical motions by the
SDSS/SEGUE, RAVE, and LAMOST surveys suggests that the local Galactic
disk is not in equilibrium in the $z$-direction.

In this paper, we considered the hypothesis that the observed bulk vertical
motions were generated by a passing satellite or dark matter subhalo.
The idea that dark matter, in one form or another, might be
responsible for heating and thickening the disk dates back to the
1980's.  \citet{lacey1985} calculated disk heating by a dark halo of
supermassive black holes while \citet{carr1987} considered dark matter
in the form of $10^6\,M_\odot$ dark clusters.  In essence,
our hypothesis is that the bulk motions seen in the data represent
the early stages of a disk heating event.
 
Our focus has been to explore the theoretical aspects of
disk-satellite interactions.  We found that the nature of the
perturbations is controlled largely by the satellite's vertical
velocity relative to the disk.  In particular, a slow moving (as
measured in the LSR) satellite induces a bending mode perturbation.
With a higher vertical velocity, higher order modes, such as the
breathing mode, are excited.  Thus, if a satellite is indeed
responsible for the bulk vertical motions seen in the solar
neighborhood, this its vertical velocity through the disk would likely
have been $\appgeq 50\,{\rm km\,s}^{-1}$.  Moreover, its surface
density would have to be comparable to that of the disk in order to
produce an appreciable perturbation.  The model satellites considered
by \citet{gomez2013} satisfy these conditions and so it is not
surprising that they found vertical perturbations in the disk that
were qualitatively similar to what was found in the data.

Single satellite simulations show that after a localized breathing
mode perturbation is produced, it is sheared by the differential
rotation of the disk.  After several orbital periods of the disk, the
perturbation assumes on a spiral-like pattern.  The situation is more
complicated with a population of satellites and it may be difficult to
disentangle initial perturbations from the accumulated long-lived
perturbations.

Our analysis, and that of \citet{weinberg1991} suggest that stars on
the tail of the energy distribution are most responsive to a breathing
mode perturbation.  Though our analyses focused on single-component
disks, it may well be that the vertical motions is a property more of
the thick disk stars, than the thin disk stars.  It is well known that
the vertical velocity dispersion and scale height are anti-correlated
with metallicity (see, for example,
\citet{bovy2012a,bovy2012b,minchev2013,minchev2014} and references
therein).  The arguments presented in this paper suggest that bulk
motions should be more prominent in the low metallicity/high $E_z$
populations.  A related issue is radial migration.
\citet{sellwood2002} argued that spiral waves can change the angular
momenta, and hence Galactocentric radii, of individual stars by $\sim
$50\%.  In analysing the cosmological simulations of disk formation by
\citet{martig2012}, \citet{minchev2013} found that satellite
interactions can also drive radial migration.  Moreover, radial
migration can bring high dispersion stars from the inner disk to the
solar neighborhood, and, as discussed above, these are the stars most
susceptible to a recent satellite interaction.

There are twenty-five known satellites of the Milky Way.  Moreover, in
a $\Lambda$CDM cosmology, the dark halo of a Milky Way-size galaxy is
expected to harbour many more nonluminous subhalos
\citep{klypin1999,moore1999}.  Thus, it is likely that the Galactic
disk has been continually perturbed over its lifetime.  In principle,
observations of bulk motions in the stellar disk could provide a probe
of the subhalo distribution.  To do so will require a suite of
simulations where the slope and amplitude of the subhalo mass function
are varied.

Over the next few years, Gaia will provide an unprecedented snapshot
of the Galaxy by making astrometric, spectral, and photometric
observations of approximately one billion Milky Way stars (See, for
example \citet{perryman2001} and \citet{bruijne2012}).  This data set
will yield a more accurate and complete map of bulk motions in the
stellar disk.  By bringing together these observations, theoretical
analysis, and N-body simulations we hope to better understand
Galactic dynamics, and in particular, interactions between the Milky
Way's disk and its satellites and dark matter subhalos.

\section*{Acknowledgements}
We thank Martin Weinberg and Martin Duncan for useful conversations.
We thank Facundo G{\'o}mez for providing the structural parameters for
their models of the Sagittarius dwarf.  LMW acknowledges the Aspen
Center of Physics for its hospitality.  This work was supported by a
Discovery Grant with the Natural Sciences and Engineering Research
Council of Canada.  MHC also acknowledges the financial support of the
Ontario Graduate Scholarship program.

\end{document}